\documentclass[a4paper,11pt]{article}
\usepackage{pos}
\usepackage{orcidlink}

\title{Progress on Sp(4) lattice theories with Grid: M\"{o}bius domain wall fermions and continuum extrapolations}
\ShortTitle{Sp(4) lattice theories with Grid: MDWFs and  continuum extrapolations}

\newcommand{\orcidauthorBENNETT}{0000-0002-1678-6701}
\newcommand{\orcidauthorLUCINI}{0000-0001-8974-8266}
\newcommand{\orcidauthorPIAI}{0000-0002-2251-0111}
\newcommand{\orcidauthorFORZANO}{0000-0003-0985-8858}
\newcommand{\orcidauthorVADACCHINO}{0000-0002-5783-5602}
\newcommand{\orcidauthorHILL}{0000-0003-2383-940X}
\newcommand{\orcidauthorHONG}{0000-0002-3923-4184}
\newcommand{\orcidauthorDELDEBBIO}{0000-0003-4246-3305}
\newcommand{\orcidauthorLIN}{0000-0003-3743-0840}
\newcommand{\orcidauthorLEE}{0000-0002-4616-2422}
\newcommand{\orcidauthorPROVATAS}{0000-0002-3132-9621}
\newcommand{\orcidauthorSIMONETTI}{0009-0002-3921-2687}

\author*[a,b,c]{Gianmarco Simonetti\,%
  \orcidlink{\orcidauthorSIMONETTI}}

\author[b,d]{Ed Bennett\,%
  \orcidlink{\orcidauthorBENNETT}}

\author[c,e]{Peter A. Boyle}

\author[c]{Luigi Del Debbio\,%
  \orcidlink{\orcidauthorDELDEBBIO}}

\author[a]{Niccolò Forzano\,%
  \orcidlink{\orcidauthorFORZANO}}

\author[c]{Ryan C. Hill\,%
  \orcidlink{\orcidauthorHILL}}

\author[f,g]{Deog Ki Hong\,%
  \orcidlink{\orcidauthorHONG}}

\author[h]{Jong-Wan Lee\,%
  \orcidlink{\orcidauthorLEE}}

\author[i,j,n]{C.-J. David Lin\,%
  \orcidlink{\orcidauthorLIN}}

\author[b,k,l]{Biagio Lucini\,%
  \orcidlink{\orcidauthorLUCINI}}

\author[a,b]{Maurizio Piai\,%
  \orcidlink{\orcidauthorPIAI}}

\author[m]{Davide Vadacchino\,%
  \orcidlink{\orcidauthorVADACCHINO}}

\author[a,b,c]{Alexis Verney-Provatas\,%
  \orcidlink{\orcidauthorPROVATAS}}

\affiliation[a]{
  Department of Physics, Faculty of Science and Engineering,
  Swansea University,\\
  Singleton Park, Swansea SA2 8PP, United Kingdom
}

\affiliation[b]{
  Centre for Quantum Fields and Gravity,
  Faculty of Science and Engineering, Swansea University,\\
  Singleton Park, Swansea SA2 8PP, United Kingdom
}

\affiliation[c]{
  School of Physics and Astronomy, The University of Edinburgh,\\
  Peter Guthrie Tait Road, Edinburgh EH9 3FD, United Kingdom
}

\affiliation[d]{
  Swansea Academy of Advanced Computing, Swansea University,\\
  Bay Campus, Fabian Way, Swansea SA1 8EN, United Kingdom
}

\affiliation[e]{
  Physics Department, Brookhaven National Laboratory,\\
  Upton, NY 11973, USA
}

\affiliation[f]{
  Department of Physics, Pusan National University,\\
  Busan 46241, Korea
}

\affiliation[g]{
  Extreme Physics Institute, Pusan National University,\\
  Busan 46241, Korea
}

\affiliation[h]{
  Particle Theory and Cosmology Group,
  Center for Theoretical Physics of the Universe,\\
  Institute for Basic Science (IBS), Daejeon 34126, Korea
}

\affiliation[i]{
  Institute of Physics, National Yang Ming Chiao Tung University,\\
  1001 Ta-Hsueh Road, Hsinchu 30010, Taiwan
}

\affiliation[j]{
  Centre for High Energy Physics, Chung-Yuan Christian University, \\
  Chung-Li 32023, Taiwan
}

\affiliation[k]{
  Department of Mathematics, Faculty of Science and Engineering,
  Swansea University,\\
  Bay Campus, Fabian Way, Swansea SA1 8EN, United Kingdom
}

\affiliation[l]{
  School of Mathematical Sciences, Queen Mary University of London,\\
  Mile End Road, London E1 4NS, United Kingdom
}

\affiliation[m]{
  Centre for Mathematical Sciences, University of Plymouth, \\
  Plymouth PL4 8AA, United Kingdom
}

\affiliation[n]{Physics Division, National Centre for Theoretical Sciences, \\
Taipei 106319, Taiwan
}

\emailAdd{s2820577@ed.ac.uk}

\abstract{
\begin{center}
\href{https://telos-collaboration.github.io}{ \includegraphics[height=1cm]{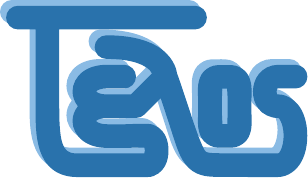}}\\
(on behalf of the TELOS collaboration)\\
\end{center}
We present preliminary results obtained in the first extensive study of the M\"{o}bius domain wall fermion (MDWF) lattice formulation of the four-dimensional \(Sp(4)\) gauge theory coupled to two dynamical fermions transforming in the fundamental representation. This theory is important for extensions of the Standard Model, as it provides the microscopic origin for both composite Higgs and dark matter models based on the \(SU(4)/Sp(4)\) coset. We scan the space of bare lattice parameters, including those entering the MDWF, and monitor spurious effects, such as the residual mass. We identify optimised regions of parameter space in which lattice artefacts affecting the global symmetries of the theory are suppressed. We then perform preliminary measurements of the lightest pseudoscalar and vector meson masses, together with the pseudoscalar decay constant. We demonstrate the potential of the MDWF formulation to enable continuum-limit extrapolations with reduced lattice discretisation effects.
}

\FullConference{The 43rd International Symposium on Lattice Field Theory (Lattice 2026)\\
July 26 to August 1, 2026\\
University of Maryland, College Park, USA\\}

\begin{document}
\maketitle

\section{Introduction}

The M\"{o}bius formulation of Domain-Wall Fermions (MDWFs)~\cite{Kaplan:1992bt,Brower:2004xi,Brower:2005qw,Brower:2012vk}, provides an  ideal lattice field theory tool for the study of the non-perturbative dynamics in proposals of new physics with composite, strongly coupled, gauge theory origin.  
The Sp(4) lattice theory coupled to $N_{\rm f}=2$ species of Dirac fermions provides a compelling realisation of composite Higgs models (CHMs)~\cite{Barnard:2013zea,Ferretti:2013kya} and strongly interacting massive particle (SIMP) dark matter models~\cite{Hochberg:2014kqa}. 
We implement the MDWF formulation of this theory in the Grid software suite~\cite{Boyle:2015tjk}, adapted to handle symplectic gauge groups~\cite{Bennett:2023gbe}, and use the Hadrons  framework~\cite{antonin_portelli_2023_8023716} to analyse it.
We report preliminary results of the MDWF algorithm optimisation, and spectroscopy measurements, that we critically compare with existing measurements~\cite{TELOS:2026alk}, obtained by implementing dynamical Wilson fermions with the use of the HiRep code~\cite{HiRepSpN}---see also Refs.~\cite{Bennett:2017kga,Lee:2018ztv,  Bennett:2019jzz, Bennett:2023rsl}. A more in-depth analysis of the observables described in this manuscript using an extended dataset will be part of the updated publication~\cite{bennett2026symplecticlatticegaugetheories} (with corresponding data and workflow~\cite{datarelease, workflowrelease}).
This and the closely related work~\cite{Bennett:2017kga,Lee:2018ztv,  Bennett:2019jzz, 
Bennett:2022yfa, 
Bennett:2022ftz, 
Bennett:2023rsl, Bennett:2023gbe, 
Bennett:2024cqv,  Bennett:2024wda, Bennett:2024tex, 
TELOS:2025ash,
TELOS:2026alk} are part of TELOS programme of exploration of ${\rm Sp}(2N)$ lattice theories.

\section{The MDWF lattice theory}

The five-dimensional Euclidean hypercubic lattice has lattice spacings $a$  in the four physical dimensions and $a_5$ along the fifth dimension. The temporal and spatial extents are $T=N_ta$ and $L=N_sa$, while the extent of the fifth dimension is $L_5=L_sa_5$. The lattice action consists of three contributions:
$
S \equiv S_g + S_f + S_{\rm PV}
$,
where $S_g$ is the gauge action, $S_f$ is the fermion action, and $S_{\rm PV}$ is the Pauli--Villars contribution. The gauge dynamics are described by the standard Wilson plaquette action for the $Sp(2N)$ gauge theory,
$
S_g \equiv
\beta\sum_x\sum_{\mu<\nu}
\left(
1-\frac{1}{2N}\operatorname{Re}\operatorname{Tr}
\mathcal{P}_{\mu\nu}(x)
\right)
$,
where the lattice coupling is $\beta\equiv 4N/g_0^2$, with $g_0$ denoting the bare gauge coupling, and $\mathcal{P}_{\mu\nu}(x)$ is the elementary plaquette at the lattice site $x$ in the $\mu\nu$ plane. The five-dimensional fermion action is
\begin{equation}
\label{eq:fermionaction}
S_f \equiv
a^4a_5
\sum_{j=1}^{N_{\rm f}}
\sum_{x,y}
\sum_{s,r=0}^{L_s-1}
\overline{\Psi}^{\,j}(x,s)
D^{\rm M}(x,s,y,r;m_j)
\Psi^j(y,r),
\end{equation}
where $\Psi^j$ is a five-dimensional fermion field of flavour $j$. In this work, we consider $N_{\rm f}=2$ degenerate fermions, with
$m_1=m_2\equiv m_0$. The fifth dimension separates the two chiral components, which are exponentially localised near its opposite boundaries. 
At finite $L_s$, the overlap between the two boundary modes produces a residual explicit breaking of chiral symmetry, which vanishes in the limit $L_s\rightarrow\infty$. The 
$D^{\rm M}$ operator entering Eq.~\eqref{eq:fermionaction} has the block-tridiagonal form~\cite{Brower:2005qw}
\begin{equation}
D^{\rm M}\equiv
\renewcommand{\arraystretch}{0.8}
\begin{bmatrix}
D_L        & D_RP_R & 0      & \cdots & -m_0D_RP_L \\
D_RP_L     & D_L    & D_RP_R & \cdots & 0          \\
0          & D_RP_L & D_L    & \cdots & 0          \\
\vdots     & \vdots & \vdots & \ddots & \vdots     \\
-m_0D_RP_R & 0      & 0      & \cdots & D_L
\end{bmatrix}\,,
\end{equation}
where $D_L=bD_{\rm W}+1$, $D_R=cD_{\rm W}-1$  and $b,c>0$ are real parameters. $D_{\rm W}$ is the four-dimensional Wilson--Dirac operator with domain-wall mass $m_5$. 
The MDWF formulation is related to the Shamir kernel~\cite{Shamir:1993zy} through

\begin{equation}
D^{\rm M}=
\alpha D_{\rm S}
=
\alpha\frac{a_5D_{\rm W}}{2+a_5D_{\rm W}},
\qquad
a_5\equiv(b-c)a>0,
\qquad
\alpha\equiv\frac{(b+c)a}{a_5}.
\end{equation}
The Shamir formulation corresponds to $\alpha=1$, but optimising the choice of $\alpha$ can reduce the value of $L_s$ required to suppress  residual symmetry breaking effects~\cite{Brower:2005qw,Brower:2012vk}. Finally, the Pauli--Villars term, $S_{\rm PV}$, introduces bosonic scalar regulator fields to cancel the unphysical heavy bulk modes of the five-dimensional formulation~\cite{Narayanan:1993sk}. 
Conventionally, we fix to $am_{\rm PV}=1$.


\section{MDWF optimisation}

\begin{figure}[t]
    \centering
    \begin{minipage}[c]{0.47\textwidth}
        \centering
        \includegraphics[width=0.9\textwidth]{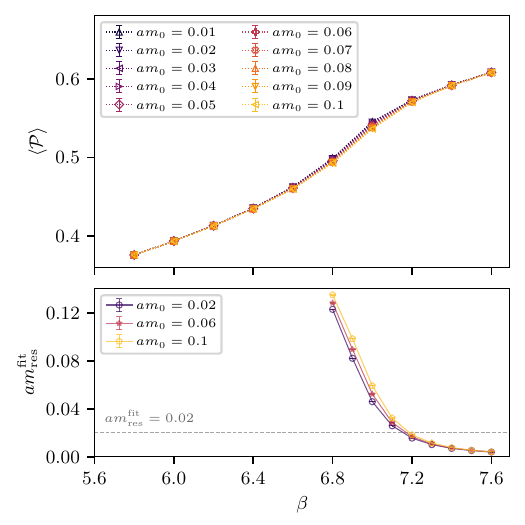}
    \end{minipage}
    \hfill
    \begin{minipage}[c]{0.47\textwidth}
        \centering
        \includegraphics[width=0.9\textwidth]{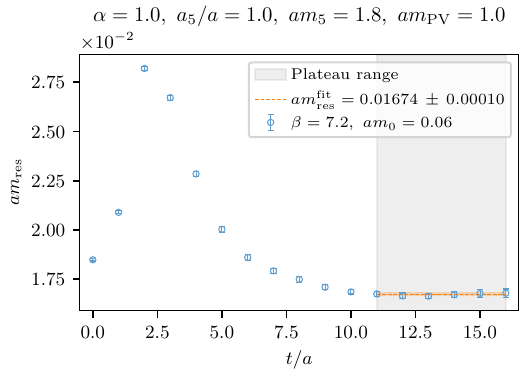}
    \end{minipage}
    \caption{Average plaquette, $\langle {\cal P}\rangle$ (top-left panel), and residual mass, $am_{\rm res}$ (bottom-left), computed in the Shamir limit of the MDWF formulation, with $\alpha=1.0$, $am_5=1.8$, $a_5/a=1.0$, and $L_s=8$, for different choices of $\beta$, and volumes $\widetilde V=8\times8^3$ and $32\times16^3$, respectively. The right panel shows a representative plateau on the folded $am_{\rm res}$, and its correlated constant fit, obtained according to Eq.~\eqref{eq:residual_mass}.}
    \label{fig:avg_plaq_mres}
\end{figure}

Our tuning procedure follows Ref.~\cite{Furman:1994ky}. The enhanced global $SU(4)$ symmetry acting on the fermions of the continuum theory is broken, hence axial-vector and pseudoscalar currents form a degenerate multiplet, $\mathbf{5}$, of the unbroken global $Sp(4)$ symmetry. It is hence sufficient to consider one single generator, $T^1$, and following the conventions of Ref.~\cite{Lee:2017uvl}, we write
\begin{equation}
\mathcal{O}^{\mathrm{AV},\,1}_{\mu}
\,=\,
\overline{Q^{1a}}\gamma_\mu\gamma_5 Q^{2a}
+
\overline{Q^{2a}}\gamma_\mu\gamma_5 Q^{1a}\,,\quad
\mathcal{O}^{1}_{\mathrm{PS}}
\,=\,
\overline{Q^{1a}}\gamma_5 Q^{2a}
+
\overline{Q^{2a}}\gamma_5 Q^{1a}.
\end{equation}
The corresponding lattice current is defined starting from the five-dimensional fermions, $\Psi(x, s)$~\cite{Furman:1994ky}:
\begin{equation}
	j_\mu^1(x, s) =  \dfrac{1}{\sqrt{2}} \left[\bar{\Psi}(x+\hat{\mu}, s)(1+\gamma_\mu) U^{\dagger}_{\mu} (x) T^1 \Psi(x, s) - \bar{\Psi}(x, s)(1-\gamma_\mu) U_{\mu} (x) T^1 \Psi(x+\hat{\mu}, s) \right] ,
\end{equation}
where $T^1 \in SU(4)/Sp(4)$ is chosen to match the conventions for $\mathcal{O}^{\mathrm{AV}, \, 1}_{\mu} $ and $\mathcal{O}^1_{\mathrm{PS}}$. 
\begin{figure}[t]
    \centering
\includegraphics[width=0.85\textwidth]{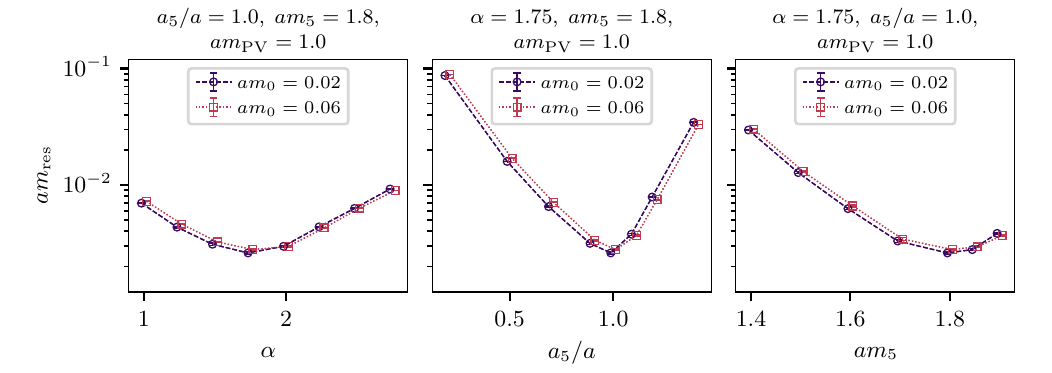}
   \caption{ \label{fig:scan}
   Representative example of MDWF parameter scans for the $Sp(4)$ theory with $N_{\rm f}=2$, showing $am_{\rm res}$ as a function of $\alpha$, $a_5/a$, and $am_5$ (left to right), varied individually. Results are obtained at $\beta=7.4$ and $am_0=0.02,\,0.06$ on lattices with $L_s=8$, $N_s=16$, and $N_t=32$.
   }
\end{figure}
The four-dimensional domain-wall axial-vector current is constructed 
as follows:
\begin{equation}
\mathcal{A}_\mu^1(x)
\equiv
\sum_{s=0}^{L_s-1}
\operatorname{sgn}\left(s-\frac{L_s-1}{2}\right)
j_\mu^1(x,s)
=
Z_A\,\mathcal{O}^{\mathrm{AV},\,1}_\mu(x)\,.
\label{eq:dw_axial_current}
\end{equation}
The second equality expresses the relation, in the limit $L_s\to\infty$, to the corresponding local axial-vector operator, via  the multiplicative renormalisation factor, $Z_A$. For a finite value of $L_s$, the incomplete separation of the left- and right-handed boundary-localised modes results in a residual breaking of the global symmetry. The corresponding axial Ward identity reads~\cite{Furman:1994ky}: 
\begin{equation}
\label{eq:pcac_dw}
\Delta_\mu\mathcal{A}^1_\mu(x)
=
2am_0\,\mathcal{O}^1_{\rm PS}(x)
+2\mathcal{O}^1_{{\rm PS},q}(x)
\simeq
2\left(am_0+am_{\rm res}\right)
\mathcal{O}^1_{\rm PS}(x)
+\mathcal{O}(a^2),
\end{equation}
where $\mathcal{O}^1_{\rm PS}$ and $\mathcal{O}^1_{{\rm PS},q}$ are the boundary and midpoint pseudoscalar densities, respectively. The residual mass, $am_{\rm res}$, quantifies the residual symmetry breaking induced by finite $L_s$~\cite{Aoki:2002vt}. We estimate it with a constant fit to the plateau, at large Euclidean time, of the ratio:
\begin{equation}
\label{eq:residual_mass}
am_{\rm res}(t)
\equiv
\displaystyle\sum_{\vec{x},\vec{y}}
\left\langle
\mathcal{O}^1_{{\rm PS},q}(\vec{y},t)
\mathcal{O}^1_{\rm PS}(\vec{x},0)
\right\rangle
{\bigg/}
\displaystyle\sum_{\vec{x},\vec{y}}
\left\langle
\mathcal{O}^1_{\rm PS}(\vec{y},t)
\mathcal{O}^1_{\rm PS}(\vec{x},0)
\right\rangle
\,.
\end{equation}

We start with the Shamir formulation ($\alpha=1$), to identify a suitable range of lattice couplings, before tuning the MDWF parameters. We fix $a_5/a=1$, $am_5=1.8$, $am_{\rm PV}=1$, and $L_s=8$, and require that the residual mass, $am_{\rm res}$, be small compared with the  bare mass, $am_0$. Figure~\ref{fig:avg_plaq_mres} shows the average plaquette, $\langle {\cal P} \rangle$, and residual mass,  $am_{\rm res}$ as functions of $\beta$, for several values of $am_0$. The plaquette varies smoothly, with no indication of a bulk phase transition, while $am_{\rm res}$ decreases with increasing $\beta$, leading  us to restrict attention to $\beta\geq 7.4$. We tune the MDWF parameters by measuring $am_{\rm res}$ while varying  $\alpha$, $a_5/a$, and $am_5$ for fixed $\beta=7.4$ and $am_{\rm PV}=1$, varying one parameter at a time.  Figure~\ref{fig:scan} exhibits marked minima: the optimisation of $\alpha$ substantially suppresses $am_{\rm res}$ relative to $\alpha=1$, and points to  preferred values of  $a_5/a=1$ and $am_5=1.8$.

\begin{figure}[t]
    \centering
    \begin{tabular}{c}
    \includegraphics[width=0.9\textwidth]{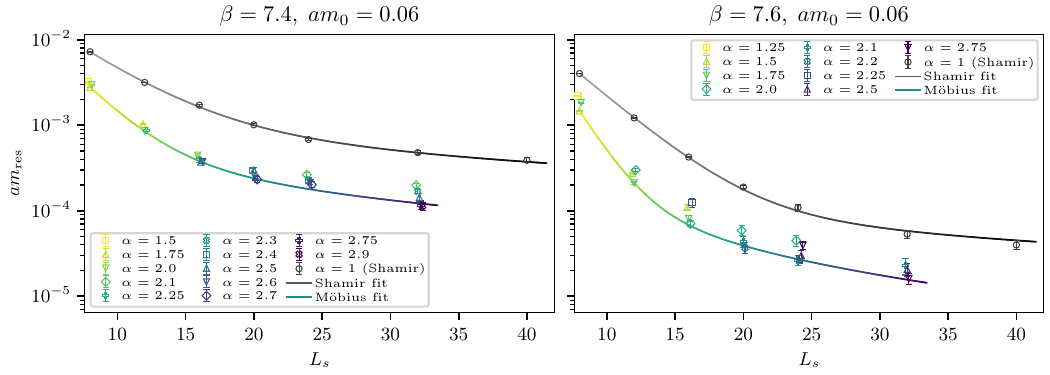} \\
    \end{tabular}
   \caption{ \label{fig:mres_Ls}
   Residual mass, $am_{\rm res}$, as a function of $L_s$ for the ${\rm Sp}(4)$ theory with $N_{\rm f}=2$, computed on lattices with extent $32\times16^3$, for fixed $am_0=0.06$, $a_5/a=1.0$, $am_5=1.8$, and $am_{\rm PV}=1.0$, and for several values of $\alpha$. The solid lines are fits to the minima in $\alpha$ using Eq.~\eqref{eq:mres_Ls}, yielding $\nu=1.25(30)$ and $1.90(32)$ for $\beta=7.4$ and $7.6$, respectively. For Shamir's fermions, $\nu=1$ is fixed.}
\end{figure}

In Fig.~\ref{fig:mres_Ls} we show $am_{\rm res}$ as a function of $L_s$ for  the choices $\beta=7.4$ and $7.6$, with fixed $am_0=0.06$, contrasting Shamir's and MDWF fermions. For MDWFs, $\alpha$ is optimised at each $L_s$, while $am_5=1.8$, $a_5/a=1.0$, and $am_{\rm PV}=1$ are fixed. In both formulations, $am_{\rm res}$ decreases with $L_s$, with the optimised MDWF results systematically below the Shamir ones. We fit the numerical results using the following approximation~\cite{RBC:2008cmd}:
\begin{equation}
\label{eq:mres_Ls}
am_{\rm res}
\simeq
c_1 e^{-\lambda_c L_s}
+
c_2 / L_s^\nu \,.
\end{equation}
The exponentially suppressed term describes the contribution of extended modes near the mobility edge, $\lambda_c$, while the power-law term accounts for low-lying localised modes. 
For the Shamir formulation, we fix $\nu=1$, while for the MDWF fits we treat $\nu$ as a free parameter, with the ideally tuned formulation expected to approach $\nu=2$~\cite{RBC:2008cmd,Brower:2012vk}. 

\section{Meson spectrum and continuum extrapolation}
\label{Sec:spectr_decayconst}

We determine the masses of the lightest flavoured pseudoscalar and vector mesons, $m_{\rm PS}$ and $m_{\rm V}$, together with the pseudoscalar decay constant, $f_{\rm PS}$, by exploiting the interpolating operators
\begin{equation}
\mathcal{O}_{\rm PS}
=
\overline{Q^1}\gamma_5 Q^2,
\qquad
\mathcal{O}_{\rm V}^{\mu}
=
\overline{Q^1}\gamma^\mu Q^2,
\qquad
\mathcal{O}_{\rm AV}^{\mu}
=
\overline{Q^1}\gamma_5\gamma^\mu Q^2,
\label{eq:meson_operators}
\end{equation}\
which receive no contributions from disconnected diagrams.
The ${\rm PS}$ and ${\rm V}$ operators source states with quantum numbers $J^P=0^-$ and $1^-$ and are used to extract $m_{\rm PS}$ and $m_{\rm V}$, respectively. The ${\rm AV}$ operator is used in combination with $\mathcal{O}_{\rm PS}$ to determine $f_{\rm PS}$. The meson masses are extracted from the large-time behaviour of the corresponding Euclidean two-point functions,
\begin{equation}
C_{XX}(t)
\equiv
\sum_{\vec{x}}
\left\langle
\mathcal{O}_X(\vec{x},t)
\mathcal{O}_X^\dagger(\vec{0},0)
\right\rangle,
\qquad
X\in\{{\rm PS},{\rm V}\}.
\label{eq:meson_correlators}
\end{equation}
At sufficiently large Euclidean times, they are dominated by the lightest state in the  channel. 

\begin{figure}[t]
    \centering
    \includegraphics[width=0.7\textwidth]{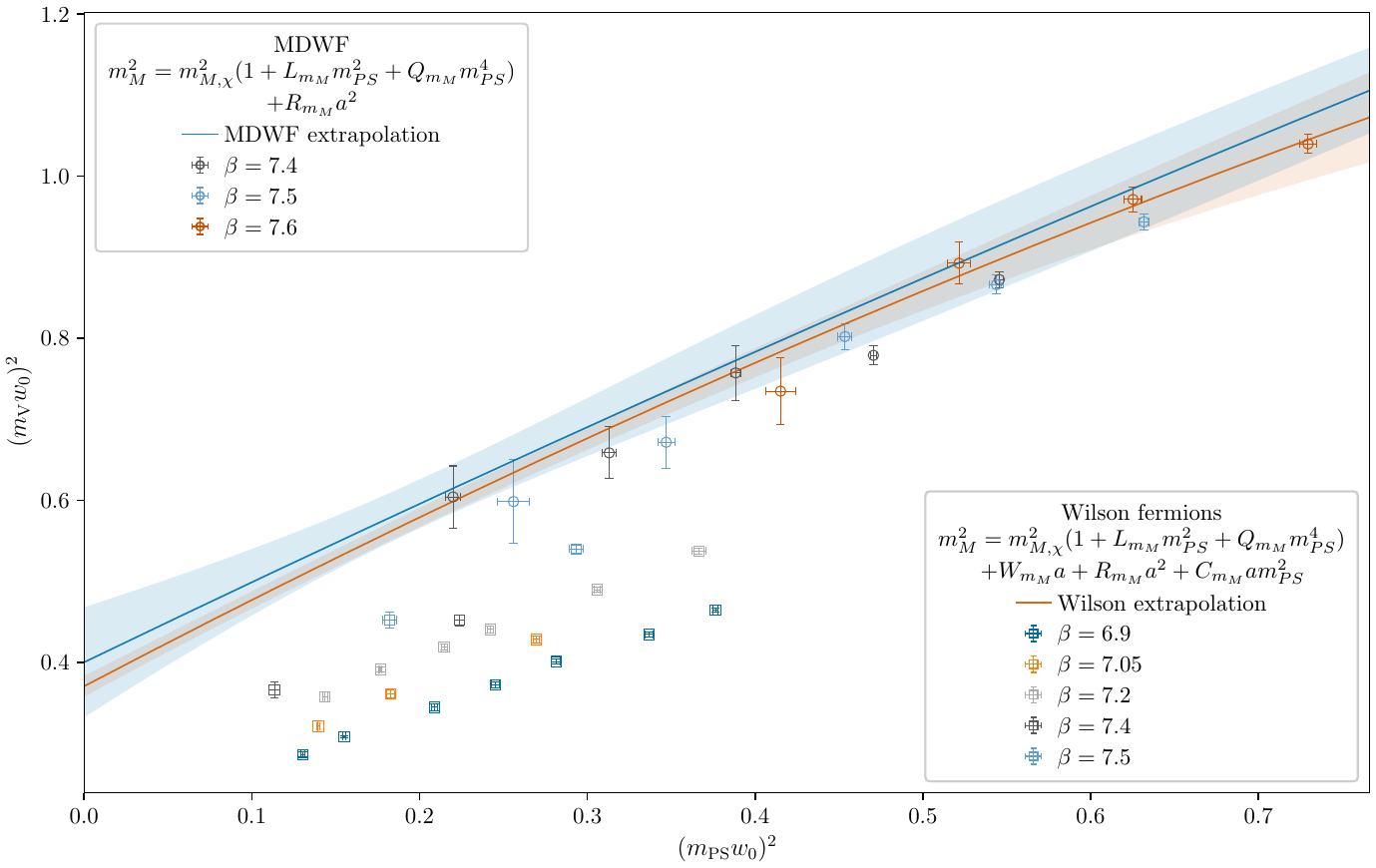}
    \caption{
    Squared vector mass,
    \(\left(m_{\rm V}w_0\right)^2\), as a function of
    \(\left(m_{\rm PS}w_0\right)^2\). Circles show preliminary MDWF results, squares denote the Wilson-fermion ones taken from Ref.~\cite{TELOS:2026alk}. The solid curves and shaded regions show the corresponding continuum extrapolations and their statistical uncertainties.
    }
    \label{fig:mv_mps}
\end{figure}

\begin{figure}[t]
    \centering
    \includegraphics[width=0.7\textwidth]{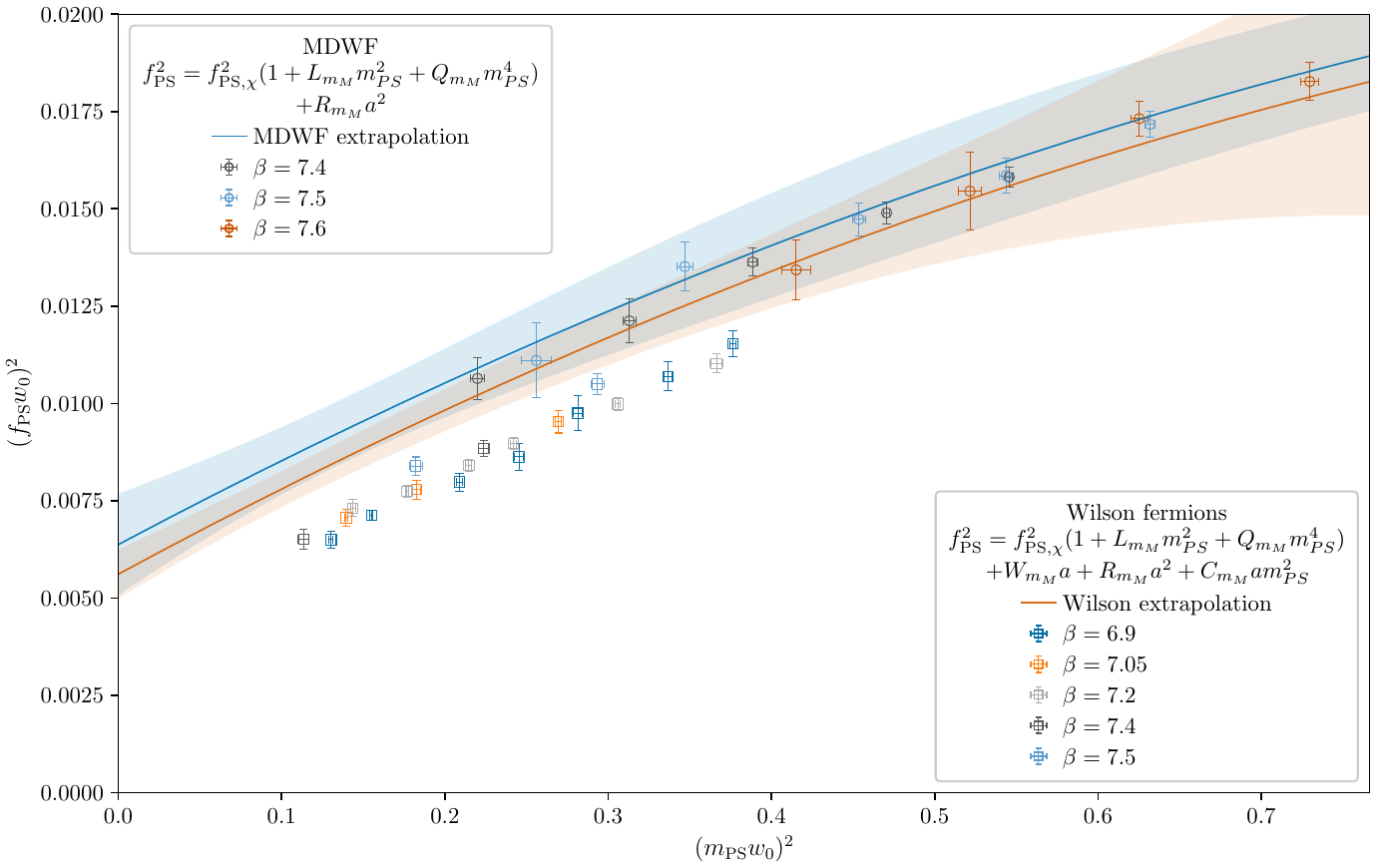}
    \caption{
    Squared renormalised pseudoscalar decay constant,
    \(\left(f_{\rm PS}^{\rm ren}w_0\right)^2\), as a function of
    \(\left(m_{\rm PS}w_0\right)^2\). Circles show preliminary MDWF results, squares denote the Wilson-fermion ones  taken from Ref.~\cite{TELOS:2026alk}. The solid curves and shaded regions show the corresponding continuum extrapolations and their statistical uncertainties.
    }
    \label{fig:fmps_mps}
\end{figure}

The pseudoscalar decay constant and its renormalised counterpart are defined by
\begin{equation}
\left\langle 0\left|
\overline{Q^1}\gamma_5\gamma_\mu Q^2
\right|{\rm PS}(p)\right\rangle
=
\sqrt{2}\,f_{\rm PS}p_\mu,
\qquad
f_{\rm PS}^{\rm ren}=Z_Af_{\rm PS},
\label{eq:fps_definition}
\end{equation}
where $Z_A$ is the renormalisation constant determined non-perturbatively~\cite{Aoki:2002vt}. We express the results in units of the Wilson-flow scale, $w_0$~\cite{Luscher:2010iy}, defined using the reference value $W_0=0.28125$~\cite{ Bennett:2022ftz}, and compare them with the Wilson-fermion results of Ref.~\cite{TELOS:2026alk}. To write the combined chiral and continuum extrapolations compactly, we introduce
the notation $\Theta\in\{m_{\rm V},f_{\rm PS}^{\rm ren}\}$ and
$x\equiv(w_0m_{\rm PS})^2$. Inspired by Wilson $ \rm W\chi PT$ and its extension to improved lattice actions~\cite{Rupak:2002sm,Symanzik:1983dc}, we perform the massless and continuum extrapolations. For the Wilson-fermion measurements, we use
\begin{align}
\left(w_0\Theta\right)^2_{\rm W}
={}&
\left(w_0\Theta^{\chi,{\rm W}}\right)^2
\left[
1+L_\Theta^{\rm W}x+Q_\Theta^{\rm W}x^2
\right]
+
W_\Theta^{\rm W}\left(\frac{a}{w_0}\right)
+
R_\Theta^{\rm W}\left(\frac{a}{w_0}\right)^2
\nonumber
+
C_\Theta^{\rm W}
\left(\frac{a}{w_0}\right)x^2\,,
\label{eq:chiPT_W}
\end{align}
while the MDWF measurements are described by
\begin{equation}
\left(w_0\Theta\right)^2_{\rm DW}
=
\left(w_0\Theta^{\chi,{\rm DW}}\right)^2
\left[
1+L_\Theta^{\rm DW}x+Q_\Theta^{\rm DW}x^2
\right]
+
W_\Theta^{\rm DW}
\left(\frac{a}{w_0}\right)^2\,.
\label{eq:chiPT_DWF}
\end{equation}
We denote as $\Theta^\chi$  the corresponding value of $\Theta$ in the massless and continuum limits. The Wilson ansatz includes both $\mathcal{O}(a)$ and $\mathcal{O}(a^2)$ corrections, together with a mixed mass-lattice-spacing term. By contrast, as long as the residual mass is sufficiently suppressed, the leading discretisation effects in the MDWF formulation appear at $\mathcal{O}(a^2)$.
Preliminary  results shown in Figs.~\ref{fig:mv_mps} and~\ref{fig:fmps_mps} demonstrate that the two continuum extrapolations agree with each other within uncertainties, providing a non-trivial validation of our implementation. At finite lattice spacing, the MDWF data lie closer to the continuum limit and show a milder dependence on $\beta$, consistently with expectations.

\section{Summary and outlook}
We presented a lattice study of the ${\rm Sp}(4)$ gauge theory with $N_{\rm f}=2$ Dirac fermions in the fundamental representation using the MDWF formulation. We first explored the MDWF parameter space by monitoring the residual mass and identified optimised choices of $\alpha$, $a_5/a$, and $am_5$ that suppress the explicit breaking of the global symmetry, at finite $L_s$. This tuning provides an efficient realisation of the domain-wall mechanism, with moderate computational cost.
We measured spectroscopy observables for the lightest accessible mesons, on ensembles with different lattice parameters. Our preliminary continuum extrapolations, compatible with previous calculations using Wilson fermions~\cite{TELOS:2026alk}, exhibit substantially milder discretisation artefacts, as expected with the MDWF formulation. These results validate our implementation and tuning strategy and provide a foundation for future studies in other regions of parameter space, and for ${\rm Sp}(4)$ theories coupled to other fermion species, previously explored with Wilson fermions~\cite{Bennett:2022yfa,Bennett:2024cqv,Bennett:2024wda,Bennett:2024tex,TELOS:2025ash}.

\begin{acknowledgments}
E.~B. is supported by the STFC Research Software Engineering Fellowship EP/V052489/1 and, in part, by UKRI via the Computational Science Centre for Research Communities (CoSeC), Grant No.~UKRI497. A.~V.-P., B.~L., E.~B., G.~S., M.~P., and N.~F. are supported by the STFC Consolidated Grant No.~ST/X000648/1. B.~L., L.~D.~D., and M.~P. received ERC funding under the European Union's Horizon 2020 research and innovation program, Grant Agreement No.~813942. L.~D.~D. is also supported by the STFC Consolidated Grant ST/T000600/1 and ST/X000494/1. N.~F. is supported by the STFC Doctoral Training Grant No.~ST/X508834/1. D.~K.~H. is supported by the Basic Science Research Program through the NRF, funded by the Ministry of Education (NRF-2017R1D1A1B06033701). J.-W.~L. is supported by IBS project IBS-R018-D1. C.-J.~D.~L. acknowledges support from NSTC Taiwan, Grant No.~112-2112-M-A49-021-MY3; the Taiwanese MoST, Grant No.~109-2112-M-009-006-MY3; and Grants No.~112-2639-M-002-006-ASP and No.~113-2119-M-007-013. B.~L. is also supported in part by the STFC Consolidated Grant No.~ST/X00063X/1. A.~V.-P. and G.~S. are supported by a studentship funded by Swansea University's Faculty of Science and Engineering and the University of Edinburgh's School of Physics and Astronomy. D.~V. is supported by the STFC Consolidated Grant No.~ST/X000680/1. This work used the DiRAC Extreme Scaling service (Tursa) at the University of Edinburgh, managed by the EPCC on behalf of the STFC DiRAC HPC Facility (www.dirac.ac.uk). The DiRAC service at Edinburgh was funded by BEIS, UKRI and STFC capital funding and STFC operations grants. DiRAC is part of the UKRI Digital Research Infrastructure. Numerical computations were performed using Supercomputing Wales SUNBIRD at Swansea University and AccelerateAI at Swansea University. Supercomputing Wales and AccelerateAI are supported by the European Regional Development Fund via Welsh Government.

{\bf Research Software and Data Availability statement}---The raw data generated in support of this work---using Grid~\cite{Boyle:2015tjk} and Hadrons~\cite{antonin_portelli_2023_8023716}---and processed data derived from it, are based on preliminary analysis. The finalised data presented in Ref.~\cite{bennett2026symplecticlatticegaugetheories} are available~\cite{datarelease, workflowrelease}.


\end{acknowledgments}

\bibliographystyle{unsrt}
\bibliography{ref}

\end{document}